\documentclass[aps,twocolumn,
superscriptaddress,
footinbib,
prl]{revtex4-1}

\usepackage{amsmath,amssymb,bm}
\usepackage{graphicx}
\usepackage{epstopdf}
\usepackage{latexsym}
\usepackage[normalem]{ulem} 
\usepackage[caption=false]{subfig}
\usepackage[usenames,dvipsnames]{color}
\usepackage{hyperref}
\usepackage{natbib}
\usepackage{xcolor}
\usepackage{physics}
\usepackage{multirow}
\usepackage{stackengine}
\usepackage{dsfont}

\usepackage{framed}

\usepackage[utf8]{inputenc} 

\begin{document}

\title{Time-Dependent Variational Principle for Open Quantum Systems with Artificial Neural Networks}

\date{\today}

\author{Moritz Reh}
\email{moritz.reh@kip.uni-heidelberg.de}
\affiliation{Kirchhoff-Institut f\"{u}r Physik, Universit\"{a}t Heidelberg, Im Neuenheimer Feld 227, 69120 Heidelberg, Germany}
\author{Markus Schmitt}
\affiliation{Institut für Theoretische Physik, Universit\"{a}t zu K\"{o}ln, 50937 K\"{o}ln, Germany}
\author{Martin G\"{a}rttner}
\affiliation{Kirchhoff-Institut f\"{u}r Physik, Universit\"{a}t Heidelberg, Im Neuenheimer Feld 227, 69120 Heidelberg, Germany}
\affiliation{Physikalisches Institut, Universit\"at Heidelberg, Im Neuenheimer Feld 226, 69120 Heidelberg, Germany}
\affiliation{Institut f\"ur Theoretische Physik, Ruprecht-Karls-Universit\"at Heidelberg, Philosophenweg 16, 69120 Heidelberg, Germany}

\begin{abstract} %\showthe\textwidth
We develop a variational approach to simulating the dynamics of open quantum many-body systems using deep autoregressive neural networks. The parameters of a compressed representation of a mixed quantum state are adapted dynamically according to the Lindblad master equation by employing a time-dependent variational principle. We illustrate our approach by solving the dissipative quantum Heisenberg model in one dimension for up to 40 spins and in two dimensions for a $4\times 4$ system and by applying it to the simulation of confinement dynamics in the presence of dissipation.
\end{abstract}

\maketitle

\textit{Introduction.} 
Solving the quantum many-body problem where it is analytically intractable constitutes a formidable challenge due to the inherent curse of dimensionality with growing system size. Today, two main routes are pursued to address this issue. On the one hand, the boundaries of classical computation are pushed by the development of tailored numerical techniques that build on the inherent structure of the quantum state of interest to find compressed representations using a subexponential number of variational parameters \cite{Schollwoeck2011, Orus2014, Cirac2020, Carrasquilla2020}.
On the other hand, recent years have brought tremendous progress in the realization of quantum simulators as originally envisioned by Feynman \cite{Feynman1982, Lloyd1996}, which emulate paradigmatic quantum many-body models using precisely controlled synthetic quantum systems of ultracold atoms in optical lattices \cite{Cooper2019, Mitra2017, Bordia2017, Schreiber2015, Bloch2008, Choi2016, Bakr2009}, trapped ions \cite{Martinez2016, Cirac1995, Gaerttner2017}, Rydberg atoms \cite{Browaeys2020,ebadi2020,scholl2020}, and many more \cite{Tsomokos2010, Hartmann2006, Ortner2009, Roumpos2007, Byrnes2008, Greentree2006, Peng2005, Mostame2008, Angelakis2007, Zhang2009, Cai2013,Choi2017timecrystal}.
These ``noisy intermediate-scale quantum`` (NISQ) simulators \cite{Preskill2018} already present a valuable expansion of our scientific toolbox, enabling the discovery of new physical phenomena \cite{Choi2016,Choi2017timecrystal,Bernien2017, Pruefer2018, Erne2018,ebadi2020,scholl2020}. In particular, they challenge the numerical state of the art and open up largely uncharted terrain, e.g., nonequilibrium quantum matter in two spatial dimensions. As the term NISQ implies, the openness of these quantum systems will play a central role for near-term applications, and accounting for it appropriately is one of the key challenges.

\begin{figure}
    \centering
    \includegraphics[width=\linewidth]{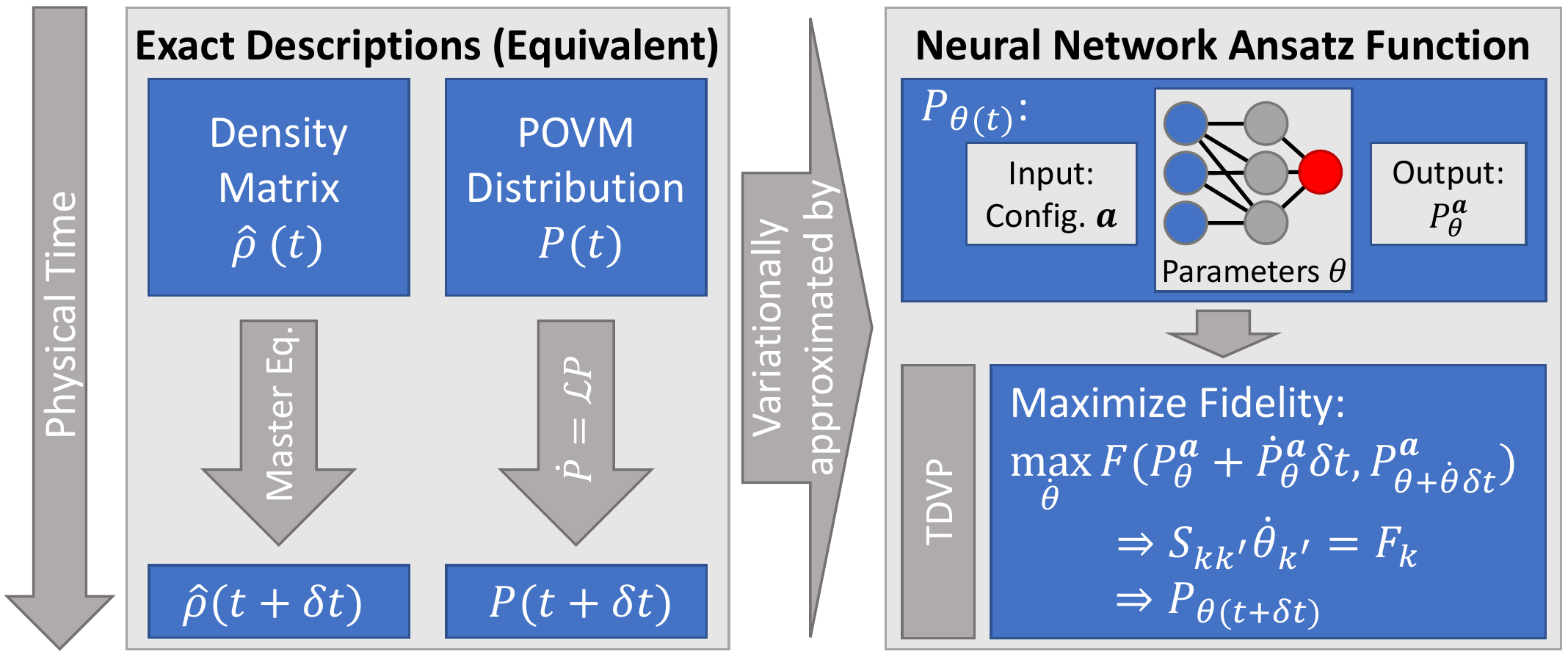}
    \caption{Illustration of the variational approach to OQS dynamics. On the left, the standard density matrix formalism is shown along with the equivalent probabilistic formulation using POVMs. 
    The right-hand side shows the variational approach, in which an artificial neural network is used as an ansatz function for the probability distribution over POVM outcomes. The main contribution of this work is the TDVP illustrated in the bottom box which leads to a general and accurate scheme for updating the network parameters according to the dynamics dictated by the master equation.
    }
    \label{fig:IntroFig}
\end{figure}

In this work, we present a novel way to simulate the dynamics of open quantum systems (OQS) using a neural network encoding of the quantum state, which is relevant for two reasons: In view of the recent experimental developments, computational tools that can keep up with the system sizes of quantum simulators also in intermediate spatial dimensions are highly desired as they allow us to certify experimental observations and provide a link to theoretical models. Simultaneously, the exploration of phenomena associated with driven dissipative systems is a major aim in itself, for which our approach opens new possibilities.

The state of an OQS is described by the density operator $\hat{\rho}$, whose dynamics, for Markovian systems, is governed by a Lindblad master equation. For a system of $N$ spin-$1/2$ particles considered here the curse of dimensionality manifests in the $4^N$ coefficients necessary to fully represent $\hat{\rho}$, which limits exact numerical treatments to small $N$.
Various numerical methods have been developed to reduce this complexity \cite{Weimer2019}, each coming with different strengths and limitations. Stochastic Monte Carlo wave function (MCWF) methods \cite{Moelmer1993, Plenio1998, Dalibard1992, Dum1992, Kornyik2019} achieve a quadratic improvement of the $N$ scaling at the cost of requiring statistical averaging. 
Semiclassical \cite{Carusotto2013, Vicentini2019a, Vicentini2018, Dagvadorj2015} and mean-field-like methods \cite{Liboff2003, Navez2010} provide a polynomial scaling in $N$ but often suffer from uncontrolled approximations and numerical instabilities.
Tensor network based approaches \cite{Schollwoeck2011, Orus2014, Wood2015, Luchnikov2019, Keever2020, Kilda2021, Werner2016, Jaschke2018, Pirvu2010, Cirac2017, Mascarenhas2015} are limited to weakly entangled states and require further approximations if applied in dimensions $d>1$ \cite{Haferkamp2020, Schuch2007, Kshetrimayum2017}.
A recently introduced class of methods, that can potentially resolve many of these issues are neural network quantum states (NQS) \cite{Carleo2017, Carrasquilla2019, Carrasquilla2019a, Schmitt2020, Luo2020, HibatAllah2020, Deng2017, Czischek2018, Neugebauer2020, Gao2017, Deng2017a, Jia2019, Zhang2018, Huang2017, Lu2019, Gan2017}. NQS have been applied successfully to OQS \cite{Hartmann2019, Yoshioka2019, Nagy2019, Vicentini2019}. A natural approach is to employ a latent state purification \cite{Torlai2018a}; 
however, this procedure has so far been restricted to shallow neural network architectures.
A more recent work uses a probabilistic representation of the quantum state \cite{Luo2020} which allows the use of deeper, more expressive networks but has the drawback of being forced to globally optimize the network parameters in each time step.

Here we introduce a numerical approach, summarized graphically in Fig.~\ref{fig:IntroFig}, that is not restricted in terms of network architectures and operates based on explicit second-order local updates, thus overcoming structural and technical limitations of previously proposed methods. 
The derivation of a first-order differential equation for the time dependence of variational parameters in the context of a probabilistic formulation of quantum mechanics is a central result of our work.
Thereby, our method expands the capabilities of previous approaches \cite{Luo2020, Hartmann2019} in terms of system sizes and timescales reached reliably. 
%The distinct difference to previous works lies in the optimization objective: Here, we work out a differential equation for the network parameters based on minimizing the difference between probability distributions as compared to minimizing a distance measure between quantum states.
This is demonstrated by the application to benchmark problems of spin systems in 1D and 2D geometries and by showing a first physics-motivated application.

\textit{Probabilistic representation.} 
Any quantum state $\hat{\rho}$ can be represented equivalently as a probability distribution $P$ over measurement outcomes using positive operator valued measures (POVMs) \cite{Peres1995, Carrasquilla2019, Carrasquilla2019a, Luo2020}: 
\begin{equation}
\label{eqn:P_from_rho}
P^\textbf{a} = \tr\left(\hat{\rho} \hat{M}^\textbf{a}\right) \,,
\end{equation}
where $\hat{M}^\textbf{a}=\hat{M}^{a_1}\otimes ..\otimes \hat{M}^{a_N}$ are measurement operators associated with the outcome $\textbf{a}=a_1..a_N$ of a tomographically complete measurement on $N$ spins. We choose $\hat{M}^i$ to be the symmetric informationally complete-POVM (SIC-POVM), or tetrahedral POVM \cite{Carrasquilla2019}. Its elements are obtained from the definition $M^a = (\mathds{1} + \vec{s}^a\cdot\vec{\sigma}) / 4$, in which the $\vec{s}^a$ form a tetrahedron on the surface of the Bloch sphere and $\vec{\sigma}$ denotes the vector of Pauli matrices. Inverting Eq.~\eqref{eqn:P_from_rho} gives
\begin{equation}
\label{eqn:rho_from_P}
\hat{\rho} = P^\textbf{a}T^{-1 \textbf{a}\textbf{a}^\prime}\hat{M}^{\textbf{a}^\prime}
\end{equation}
with the overlap matrix $T^{\textbf{a}\textbf{a}^\prime} = \tr\left(\hat{M}^{\textbf{a}}\hat{M}^{\textbf{a}^\prime}\right)$, where implicit summation over repeated indices is assumed from here on.
Since the POVM elements $\hat{M}^\textbf{a}$ form an operator basis, observables can be decomposed as $\hat{O}=\Omega^\textbf{a}\hat{M}^\textbf{a}$ and their expectation values become $\langle \hat O \rangle = P^{\textbf{a}}\Omega^{\textbf{a}}$.
Compared to the complex-valued density matrix or its purification, the probabilistic representation has the advantage that it allows us to directly leverage the highly sophisticated toolbox for generative models developed in recent years by the machine learning (ML) community \cite{Rumelhart1986, Hochreiter1997, vaswani2017attention}.

The dynamics of Markovian OQS is described by the Lindblad master equation \cite{Weimer2019}
\begin{equation}
\label{eqn:Lindbladian_rho}
\dot{\hat{\rho}} = -i[\hat{H}, \hat{\rho}] + \gamma \sum_i \left(\hat{L}^i\hat{\rho} \hat{L}^{i^{\dagger}} - \frac{1}{2} \left\lbrace \hat{L}^{i^{\dagger}} \hat{L}^i, \hat{\rho} \right\rbrace\right)
\end{equation}
with $[.,.]$ ($\lbrace\cdot ,\cdot\rbrace$) denoting the (anti-)commutator. The operators $\hat{L}^i$ are commonly referred to as jump operators and are representative of the dissipative processes that the system is subject to.
Differentiating Eq.~(\ref{eqn:P_from_rho}) and inserting Eqs.~\eqref{eqn:Lindbladian_rho} and (\ref{eqn:rho_from_P}) allows to state the master equation in the probabilistic formulation: 
\begin{equation}
\label{eqn:Lindbladian_P}
\dot{P}^{\textbf{a}} = \mathcal{L}^{\textbf{ab}}P^\textbf{b} \,.
\end{equation}
The full expression for the Lindbladian $\mathcal{L}$ is given in the Supplemental Material \cite{SM}.
Crucially, $\mathcal{L}$ is sparse since the restriction to one- and two-body interactions in Eq.~(\ref{eqn:Lindbladian_rho}) is preserved in the probabilistic reformulation, allowing us to evaluate its action efficiently.

Nonetheless, Eq.~\eqref{eqn:Lindbladian_P} is numerically intractable for many-body systems, because of the exponentially large number of coefficients $P^{\mathbf a}$. In the following we employ a variational approximation by introducing a trial distribution $P^{\mathbf a}_\theta$ with variational parameters $\theta$, see Fig.~\ref{fig:IntroFig}. The compressed representation of the state in a polynomial number of variational parameters renders the approach numerically feasible.

\begin{figure*}[t]
    \centering
    \includegraphics[width=\linewidth]{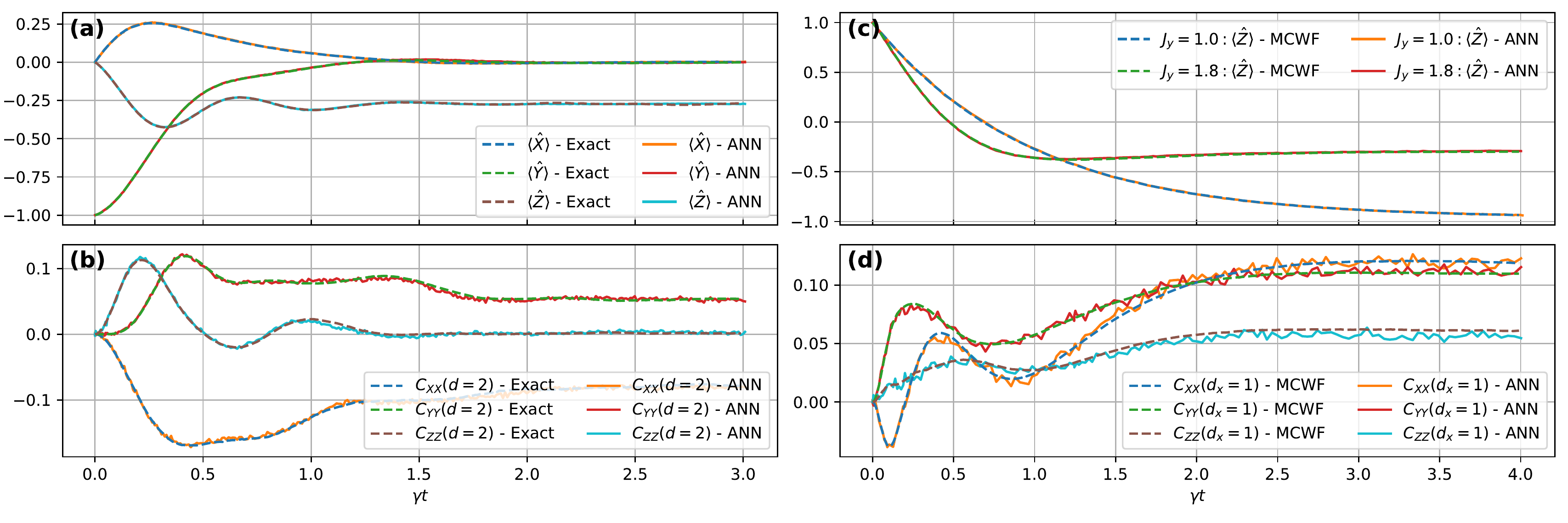}
    \caption{(a) and (b) Mean magnetizations and next-nearest neighbor connected correlation functions [e.g.,\ $C_{XX}(d=2)=\sum_i\langle \hat X_i \hat X_{i+2}\rangle^c/N$] as a function of time in the anisotropic 1D Heisenberg model for $N=40$ spins starting in the product state $\langle \hat{Y}\rangle = -1$. Nearest neighbor couplings are given by $\vec{J}/\gamma =(2, 0, 1)$, $h_z/\gamma = 1$ and the dissipation channel is $\hat{L}=\hat{\sigma}^- =\frac{1}{2}(\hat{X}-i\hat{Y})$. The exact data are obtained for $N=10$ spins. (c) and (d) Mean $z$ magnetizations and nearest neighbor connected correlation functions (for $J_y/\gamma=1.8$) in a $4\times 4$ anisotropic 2D Heisenberg lattice with nearest neighbor couplings $\vec{J}/\gamma = [0.9, 1.0 (1.8), 1.0]$ and the same decay as in (a) and (b), starting in the product state $\langle \hat{Z} \rangle = 1$.}
    \label{fig:Carra_Comparison}
\end{figure*}
\textit{Time dependent variational principle for POVMs.} 
The main theoretical contribution of our work is a time dependent variational principle (TDVP) for POVM-probability distributions which dictates the time dependence of the network parameters $\theta(t)$ by determining the closest approximation of the Lindbladian dynamics within the variational manifold. 
The starting point is a distance measure $\mathcal D(P,Q)$ for probability distributions $P,Q$. Assuming a small time step $\tau$ at time $t$ with associated network parameters $\theta(t)$, the aim is to minimize the distance between the updated POVM-probability distribution $P^{\mathbf a}_{\theta(t)+\dot\theta\tau}$ and the time-propagated distribution $P^{\mathbf a}_{\theta(t)}+\tau\mathcal L^{\mathbf{ab}}P^{\mathbf b}_{\theta(t)}$ that follows from Eq.~\eqref{eqn:Lindbladian_P}. We found that two natural choices for the distance measure $\mathcal D$ are equivalent for this purpose, because they describe locally identical geometries. The first is the Hellinger distance $\mathcal D_H(P,Q)=1-F(P,Q)$, which is defined via the Bhattacharyya coefficient (or classical fidelity) $F(P,Q)=\sum_{\mathbf a}\sqrt{P^{\mathbf a}Q^{\mathbf a}}$. The second is the Kullback-Leibler divergence $\mathcal D_{KL}(P,Q)=\sum_{\mathbf a}P^{\mathbf a}\log\big(P^{\mathbf a}/Q^{\mathbf a}\big)$.
In both cases, a second order consistent small-$\tau$ expansion of $\mathcal D_{H/KL}\big(P^{\mathbf a}_{\theta(t)+\dot\theta\tau},P^{\mathbf a}_{\theta(t)}+\tau\mathcal L^{\mathbf{ab}}P^{\mathbf b}_{\theta(t)}\big)$ and subsequently demanding stationarity to find the optimal parameter update $\dot\theta$ yields the TDVP equation
\begin{equation}
\label{eqn:tdvp_final}
S_{kk^\prime} \dot{\theta}_{k^\prime} = F_k\ .
\end{equation}
Here, $S$ denotes the Fisher-metric $S_{kk^\prime} = \langle O^\textbf{a}_k O^\textbf{a}_{k^\prime}\rangle^c_{\textbf{a}\sim P}$, $F_k=\langle O^\textbf{a}_k\mathcal{L}^\textbf{ab}\frac{P^\textbf{b}}{P^\textbf{a}}\rangle^c_{\textbf{a}\sim P}$ and the repeated indices $\textbf{a}$ inside the brackets are not summed over. The brackets denote connected correlation functions $\langle AB \rangle^c = \langle AB \rangle - \langle A \rangle\langle B \rangle$ of expectation values with respect to the POVM-distribution $P$ and $O^\textbf{a}_k =\partial_{\theta_k} \log P^\textbf{a}$. We provide a detailed derivation of Eq.~\eqref{eqn:tdvp_final} in the Supplemental Material \cite{SM}.
It is worth noting that for models that are normalized by default [such as Recurrent Neural Networks (RNNs)] $\langle O^\textbf{a}_k\rangle = 0$. Usually, Eq.~(\ref{eqn:tdvp_final}) is ill conditioned and needs to be regularized. Here, advanced regularization schemes such as described in Ref. \cite{Schmitt2020} are applicable but they did not turn out to be crucial for the test cases we consider.

The TDVP equation \eqref{eqn:tdvp_final} exhibits a number of features beneficial for the numerical time evolution of $\theta(t)$. As a result of the employed short-time expansion, the variational optimization problem becomes convex and information about the local geometry of the variational manifold is taken into account in the form of the Fisher-metric $S$. Upon inverting $S$, the differential equation can be solved straightforwardly with explicit integration schemes. Importantly, Monte Carlo sampling is only required once per time step. These features are in contrast to the implicit integration scheme presented in Ref.~\cite{Luo2020}, where Monte Carlo sampling is required at each optimization step performed for the iterative global minimization of a nonconvex cost function.

\textit{Network architecture.} Neural networks are highly nonlinear universal function approximators in the limit of large networks \cite{Cybenko1989, Hornik1991, Pinkus1999}. 
For the purpose of generative modeling autoregressive networks are advantageous, because they enable direct generation of uncorrelated samples; therefore, various autoregressive architectures have recently been explored for NQS \cite{Sharir2020,HibatAllah2020,Luo2020,Luo2021,Lin2021}. In the following, we employ RNNs, which belong to this family of network architectures (see Supplemental Material \cite{SM} for details).
%Among the many different neural network architectures, RNNs are particularly well suited to be used as NQS, given their property of a tractable likelihood, which, e.g., allows for exact sampling \cite{HibatAllah2020}. RNNs fall into the larger family of autoregressive networks, which encode probability distributions by partitioning them into products of conditional probability $P^\textbf{a}=\prod_i P(a_i|a_{<i})$. 
% For a more detailed discussion on the utilized approach we refer the reader to the SM \cite{SM}.

\textit{Numerical results.} 
To illustrate the accuracy and scalability of our method we apply it to the anisotropic Heisenberg model 
\begin{equation}
\hat H = \sum_{\langle ij\rangle} \left( J_x \hat X_i \hat X_j+ J_y \hat Y_i \hat Y_j+ J_z \hat Z_i \hat Z_j \right) + \sum_i h_z \hat Z_i
\end{equation}
with nearest neighbor interactions and periodic boundary conditions, which was also used in Ref. \cite{Luo2020} as a benchmark system.
The considered decoherence channel is spontaneous decay given by the jump operator $\hat L=\hat \sigma^-=(\hat X-i\hat Y)/2$ acting on each spin.
% Figures~\ref{fig:Carra_Comparison}(a) and (b) show the dynamics of magnetization and next-nearest neighbor correlations for a chain of $N=40$ spins initialized in a product state with $\langle \hat Y \rangle=-1$ for all spins. The nearly perfect agreement with an exact integration of Eq.~\eqref{eqn:Lindbladian_rho} for $N=10$ spins (dashed lines) shows that our method is stable and accurate and converges to the correct steady state at late times. Finite size effects are already negligible for $N=10$, allowing to use this as a reference. Small deviations are observed for the correlations. Possible reasons for this are limited expressivity of the chosen network, or an insufficient fine tuning of the simulation hyperparameters, such as the number of samples that are used to estimate $S$ and $F$ in Eq.~(\ref{eqn:tdvp_final}) ($10^4$ in this case).
We obtain benchmark data using exact simulations for $N=10$ spins and test our approach in the case of $N=10$ \cite{SM} and $N=40$ [Figs.~\ref{fig:Carra_Comparison}(a) and (b)] spins, where we compare magnetizations and next-nearest neighbor correlators. Since finite-size effects are negligible to good approximation for systems with more than $N=10$ spins, we can use the exact data for $N=10$ spins as comparison for the case of $N=40$ spins studied in the main text.
The noise in the correlation signal is due to the finite number of samples that are used to evaluate the observables. One observes slight deviations in the correlation functions, which may be attributed to both an imperfect choice of hyperparameters and the stochastic nature of the proposed method.
We found that the sample size needed to reach a given precision does not need to be increased when transitioning to larger systems, as the overall noise decreases thanks to a self-averaging effect in the translationally invariant system. 

Figures~\ref{fig:Carra_Comparison}(c) and~\ref{fig:Carra_Comparison}(d) show results for a $4\times 4$ lattice initialized in a product state with $\langle \hat Z \rangle=1$. In panel (c) we compare the magnetization for two different parameter choices to results from MCWF stochastic integration with 500 trajectories, showing perfect agreement. Exact integration of Eq.~(\ref{eqn:Lindbladian_rho}) would be exceedingly costly in this case. We provide a comparison of a $3\times 3$ lattice to exact dynamics in the Supplemental Material \cite{SM}, which shall serve as a numerically exact benchmark. Nearest neighbor correlations shown in panel (d) for the case of $J_y/\gamma=1.8$ show small deviations at late times which we attribute to the finite number of samples used for estimating the updates $\dot{\theta}$ \cite{SM}.

\begin{figure}[t!]
    \centering
    \includegraphics[width=\linewidth]{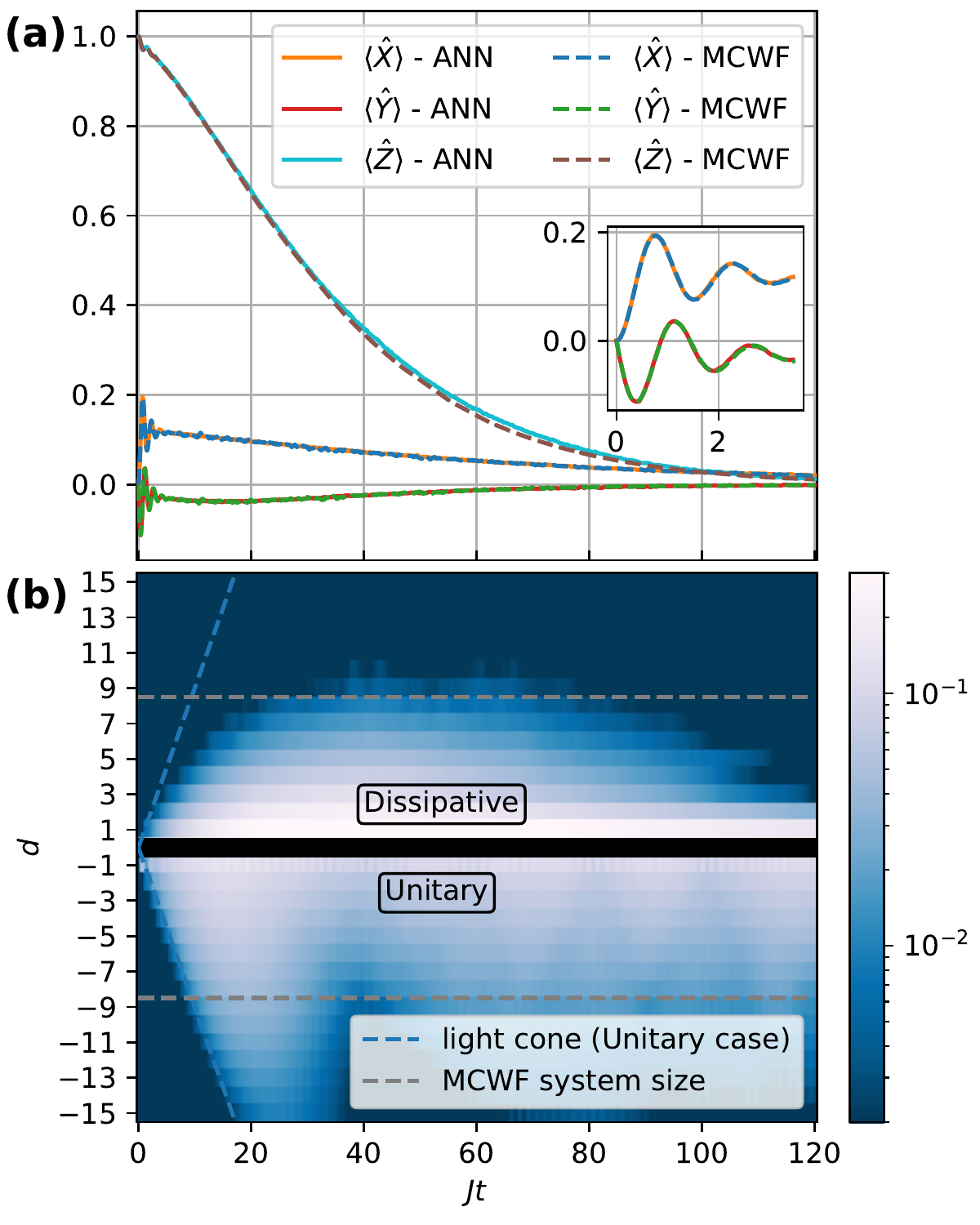}
    \caption{(a) Mean magnetizations in a spin chain of length $N=32$ with the quench parameters $h_x/J_z=0.25$, $h_z/J_z=0.05$ and the dissipation channel $\hat L=\hat{Z}$ with relative strength $\gamma / J_z = 0.25$ compared to MCWF-data for $N=16$ spins starting in the product state $\langle \hat Z\rangle = 1$. (b) Spreading of correlations in the spin chain. Top: Dissipative system with $\gamma=0.25$, Bottom: MPS simulation of the unitary system where $\gamma=0.0$. After an initial linear light-cone spreading, the nature of the dissipative propagation grows more diffusive, before all correlations eventually vanish. Notice that the slight deviations in panel (a) coincide with the time at which the dissipative correlations cross the MCWF system size boundary.}
    \label{fig:Confinement}
\end{figure}

Having benchmarked our approach on generic spin models, we now apply it to a  physical scenario to gauge the effect of decoherene for large problem instances. 
It was recently shown  that confinement dynamics, as found for quarks in quantum chromodynamics, can be realized in the Ising spin model with transverse and longitudinal fields \cite{Kormos2016}
\begin{equation}
\hat H = \sum_{\langle ij\rangle}  J_z \hat Z_i \hat Z_j + \sum_i \left( h_z \hat Z_i + h_x \hat X_i \right) \,.
\end{equation}
In this system pairs of domain walls form after a quench. For $h_z=0$ these domain walls can propagate freely while for  finite $h_z$ the separation between them comes with an energy cost leading to confinement. This phenomenon manifests in a buildup of dominant spin-spin correlations that is limited to short distances and a much weaker light-cone spreading due to the propagation of bound domain-wall pairs. Signatures of this effect have been observed recently for moderate system sizes on IBMQ \cite{Vovrosh2020}. 

Here we study in what way dissipation influences the signature spreading of spatio-temporal correlations. We consider single particle dephasing  $\hat L=\hat{Z}$ as the dissipation channel, significantly altering the nature of the spreading on timescales $\gamma t\gtrsim 1$.

Figure~\ref{fig:Confinement} shows results for a typical scenario with a dephasing rate of $\gamma = 0.25J$ and $N=32$.
The magnetization [panel (a)] initially shows coherent oscillations (inset) which are quickly damped out followed by a slow relaxation of all magnetizations towards zero. The dashed lines show MCWF simulations for $N=16$.

The top half of panel (b) shows the correlation dynamics in the considered dissipative scenario using the hitherto described numerical approach. For comparison, we show the corresponding unitary dynamics simulated using matrix product states \cite{Schollwoeck2011, Orus2014} (MPS) on the bottom half of panel (b). In the unitary case correlations initially show a light cone spreading. In contrast, the dissipative dynamics deviates from this light cone even for short times as the dissipation results in correlation growth that we find to be consistent with diffusive spreading on intermediate timescales \cite{SM}. At long times all correlations decay and the system approaches the featureless steady state $\hat \rho(t\to\infty)\propto \mathds{1}$.

The ability to simulate these dynamics is a direct consequence of the polynomial scaling of the described ansatz. The system size of the MCWF-approach ($N=16$) is plotted in panel (b) as a dashed gray line. As is obvious from the chosen color-scale cutoff the MCWF-approach suffers from finite-size effects at around $Jt=40$ when correlations beyond $d=8$ build up. This is also the time at which the $z$ magnetization in panel (a) deviates, suggesting that this deviation is due to finite-size effects present in the MCWF simulation.

\textit{Discussion and outlook.} We have introduced a novel method that allows the variational simulation of open quantum dynamics based on the efficient encoding of the quantum state in an artificial neural network and demonstrated its potential. Our method scales approximately cubically with the number of spins $N$, since the computationally intense part is obtaining $F$ in Eq.~\eqref{eqn:tdvp_final}. For that purpose, the nonvanishing entries of $\mathcal L^{\mathbf{ab}}$, the cost of evaluating a plain RNN, and the incorporation of translational symmetry each contribute a factor of $N$ to the computational cost. Importantly, however, the algorithm admits massive parallelization, e.g., on GPU clusters \cite{Schmitt2020,yang2020}, which allows for extensive control of the total compute wall time. Here, different levels of parallelization are exploited: The algorithm permits treating the samples independently from each other, allowing us to employ multiple accelerators which communicate via MPI. At the same time, the batched network evaluation of the configurations allows for convenient vectorization of the computations on the single accelerator level.

For future research it will be crucial to better understand the limitations of finite network architectures to represent physically relevant quantum states.
An obvious shortcoming of the probabilistic state representation is that the positivity of the density operator is not guaranteed; understanding the consequences will be key for further progress.

The presented combination of the probabilistic formulation of mixed quantum states with a TDVP opens new possibilities for the investigation of driven-dissipative many-body systems in regimes that are challenging for other approaches, for example, to study two-dimensional systems or the propagation of information across large distances \cite{Kastoryano2013, Sweke2019}. Future work employing the TDVP for OQS could address the emergence of glassy dynamics \cite{Poletti2013, Kucsko2018, Choi2017, Everest2017} or self-organization in OQS \cite{Zhu2015, Helmrich2020}.  Furthermore, the developed technique is not restricted to solving the Lindblad equation; it could be generalized for other use cases of master equations with large discrete configuration space, such as disease dynamics models \cite{Jenkinson2012, Keeling2008} or the chemical master equation \cite{Wolf2010}.

\acknowledgments
The MCWF data was obtained using QuTip \cite{Johansson2012}.
The TDVP algorithm was implemented using the jVMC codebase \cite{jvmc} and the JAX library \cite{Bradbury2018}. 
We acknowledge contributions by Thomas Gasenzer and Felix Behrens at early stages of this work. We thank Johannes Schachenmayer for support with MPS simulations.
This work is supported by the Deutsche Forschungsgemeinschaft (DFG, German Research Foundation) under Germany’s Excellence Strategy EXC2181/1-390900948 (the Heidelberg STRUCTURES Excellence Cluster) and within the Collaborative Research Center SFB1225 (ISOQUANT). This work was partially financed by the Baden-Württemberg Stiftung gGmbH. The authors acknowledge support by the state of Baden-Württemberg through bwHPC
and the German Research Foundation (DFG) through Grant No INST 40/575-1 FUGG (JUSTUS 2 cluster). The authors gratefully acknowledge the Gauss Centre for Supercomputing e.V. (www.gauss-centre.eu) for funding this project by providing computing time through the John von Neumann Institute for Computing (NIC) on the GCS Supercomputer JUWELS \cite{JUWELS} at Jülich Supercomputing Centre (JSC).

% \bibliography{refs}
% % \printbibliography


% appendix needs to be in an extra file for PRL
\clearpage
\newpage
\appendix

\begin{center}
    \textbf{Supplementary Materials}
\end{center}

\section{Derivation of the TDVP equation}
The basic idea of a TDVP is to minimize the distance 
\begin{equation}
\mathcal{D}\left(P_{\theta(t)} + \dot{P}_{\theta(t)} \tau, P_{\theta(t)} + \sum_k \frac{\partial P_{\theta(t)}}{\partial \theta_k}\dot{\theta}_k \tau\right),
\end{equation}
between the evolved state at time $t+\tau$ and the network with a set of yet unknown update parameters $\dot{\theta}$  at each time $t$.
Here, we exemplarily derive Eq.~\eqref{eqn:tdvp_final} from the Hellinger distance $\mathcal{D}_H(P, Q)$, i.e.\ by maximizing the classical fidelity $F(P,Q)=1-\mathcal{D}_H(P, Q)$. As noted in the main text, an equivalent derivation is possible using the Kullback-Leibler divergence $\mathcal{D}_{KL}$.
For better readability, we drop the time index and continue with the optimality condition
\begin{equation}
\begin{aligned}
0 &= \frac{\partial}{\partial \dot{\theta}_k} F\left(P + \dot{P}\tau, P + \sum_{k^\prime}\frac{\partial P}{\partial \theta_{k^\prime}}\dot{\theta}_{k^\prime}\tau\right)\\
&=\frac{\partial}{\partial \dot{\theta}_k}\sum_{\textbf{a}} P^\textbf{a} \sqrt{1 + a\tau + b \tau^2},
\end{aligned}
\end{equation}
where $a$ and $b$ are given by
\begin{equation}
\begin{aligned}
a &= \frac{\partial \log P^\textbf{a}}{\partial t} + \sum_{k^\prime} \frac{\partial  \log P^\textbf{a}}{\partial \theta_{k^\prime}}\dot{\theta}_{k^\prime},\\
b &= \frac{\partial \log P^\textbf{a}}{\partial t}\sum_{k^\prime}\frac{\partial \log P^\textbf{a}}{\partial \theta_{k^\prime}}\dot{\theta}_{k^\prime}.
\end{aligned}
\end{equation}
Next we perform a second order expansion of the square root in the time step $\tau$:
\begin{equation}
\sqrt{1 + a\tau + b\tau^2} = 1 + \frac{a\tau}{2} + \frac{\tau^2}{8} (4b - a^2) + \mathcal{O}(\tau^3).
\end{equation}
Using that the normalization of $P$ is conserved under the time evolution one finds that the term linear in $\tau$ vanishes:
\begin{equation}
\begin{aligned}
P^\textbf{a}a&=\sum_\textbf{a}\left( \dot{P}^\textbf{a} + \sum_{k^\prime} \frac{\partial P^\textbf{a}}{\partial \theta_{k^\prime}}\dot{\theta}_{k^\prime}\right)\\
&=\sum_{k^\prime} \dot{\theta}_{k^\prime}\frac{\partial}{\partial \theta_{k^\prime}}\sum_\textbf{a} P^\textbf{a}\\
&=\sum_{k^\prime}\dot{\theta}_{k^\prime}\frac{\partial}{\partial \theta_{k^\prime}}1\\
&=0.
\end{aligned}
\label{eqn:TDVP_linear_term_vanishes}
\end{equation}
Thus, the optimality condition becomes
\begin{equation}
\begin{aligned}
0 &=\frac{\partial}{\partial \dot{\theta}_k}\sum_\textbf{a} \frac{P^\textbf{a}}{P^{\textbf{a}^2}}\left(4\dot{P}^\textbf{a}\sum_{k^\prime}\frac{\partial P^\textbf{a}}{\partial \theta_{k^\prime}}\dot{\theta}_{k^\prime}-(\dot{P}^\textbf{a} + \sum_{k^\prime}\frac{\partial P^\textbf{a}}{\partial \theta_{k^\prime}}\dot{\theta}_{k^\prime})^2\right)\\
&=-\frac{\partial}{\partial \dot{\theta}_k}\sum_\textbf{a} \frac{P^\textbf{a}}{P^{\textbf{a}^2}}\left(\dot{P}^\textbf{a} -\sum_{k^\prime} \frac{\partial P^\textbf{a}}{\partial \theta_{k^\prime}}\dot{\theta}_{k^\prime}\right)^2\\
&=-\frac{\partial}{\partial \dot{\theta}_k}\sum_\textbf{a} P^\textbf{a}\left(\frac{\partial \log P^\textbf{a}}{\partial t} - \sum_{k^\prime}\frac{\partial \log P^\textbf{a}}{\partial \theta_{k^\prime}}\dot{\theta}_{k^\prime}\right)^2\\
&=2\sum_\textbf{a} P^\textbf{a}\frac{\log P^\textbf{a}}{\partial \theta_k}\left(\frac{\partial \log P^\textbf{a}}{\partial t} - \sum_{k^\prime}\frac{\partial \log P^\textbf{a}}{\partial \theta_{k^\prime}}\dot{\theta}_{k^\prime}\right).
\label{eqn:TDVP_derivation_penultimatestep}
\end{aligned}
\end{equation}
Dropping the factor of 2 we obtain an equation for the optimal parameter update $\dot{\theta}$:
\begin{equation}
\begin{aligned}
0 =& \underbrace{\sum_\textbf{a} P^\textbf{a}\frac{\partial \log P^\textbf{a}}{\partial t} \frac{\partial \log P^\textbf{a}}{\partial \theta_k}}_{=F_k}\\
&-\sum_{k^\prime}\underbrace{\sum_\textbf{a} P^\textbf{a} \frac{\partial \log P^\textbf{a}}{\partial \theta_k}\frac{\partial \log P^\textbf{a}}{\partial \theta_{k^\prime}}}_{=S_{kk^\prime}}\dot{\theta}_{k^\prime}.\\
\end{aligned}
\end{equation}
Importantly we can now tackle the sum over the exponentially many indices $\textbf{a}$ by sampling according to the encoded probabilities $P^\textbf{a}$ since both $F$ and $S$ are proportional to $P^\textbf{a}$. This is a unique property of $\mathcal{D}_H$ and $\mathcal{D}_{KL}$ 
while other distance measures, as for example the $L^2$ norm, do not lead to expressions of a form that can be efficiently evaluated from Monte Carlo samples.
Further, inserting the probabilistic form of the Lindblad master equation leads to 
\begin{equation}
\begin{aligned}
F_k &= \sum_\textbf{a} P^\textbf{a}\frac{\partial \log P^\textbf{a}}{\partial t} \frac{\partial \log P^\textbf{a}}{\partial \theta_k} \\
&=\left\langle \mathcal{L}^{\textbf{ab}}\frac{P^\textbf{b}}{P^\textbf{a}}\frac{\partial \log{P^\textbf{a}}}{\partial \theta_k}\right\rangle_{\textbf{a}\sim P}\\
\end{aligned}
\end{equation}
and 
\begin{equation}
\begin{aligned}
S_{kk^\prime} &= \sum_\textbf{a} P^\textbf{a} \frac{\partial \log{P^\textbf{a}}}{\partial \theta_k}\frac{\partial \log{P^\textbf{a}}}{\partial \theta_{k^\prime}}\\
&= \left\langle \frac{\partial \log{P^\textbf{a}}}{\partial \theta_k}\frac{\partial \log{P^\textbf{a}}}{\partial \theta_{k^\prime}} \right\rangle_{\textbf{a}\sim P} \,.
\end{aligned}
\end{equation}
The same derivation can be carried out without assuming normalization. In this case the form of $S$ and $F$ is altered to
\begin{equation}
\begin{aligned}
P^\textbf{a}&\rightarrow\frac{P^\textbf{a}}{\sum_{\textbf{b}}P^\textbf{b}}\\
\log P^\textbf{a}&\rightarrow\log P^\textbf{a} - \log \sum_{\textbf{b}}P^\textbf{b}\\
\frac{\partial \log P^\textbf{a}}{\partial \theta_k}&\rightarrow\frac{\partial \log P^\textbf{a}}{\partial \theta_k}-\left\langle\frac{\partial \log P^\textbf{a}}{\partial \theta_k}\right\rangle_{\textbf{a}\sim P}\\
\frac{\partial \log P^\textbf{a}}{\partial t}&\rightarrow\frac{\partial \log P^\textbf{a}}{\partial t}-\left\langle\frac{\partial \log P^\textbf{a}}{\partial t}\right\rangle_{\textbf{a}\sim P} \,,
\end{aligned}
\end{equation}
where the last two lines are obtained using
\begin{equation}
\begin{aligned}
&\frac{\partial}{\partial \theta_k} \left( \log P^\textbf{a} - \log \sum_\textbf{b} P^\textbf{b} \right)\\
=&\frac{\partial\log P^\textbf{a}}{\partial \theta_k} - \frac{\sum_\textbf{b}\frac{\partial P^\textbf{b}}{\partial \theta_k}}{\sum_\textbf{c}P^\textbf{c}}\\
=&\frac{\partial\log P^\textbf{a}}{\partial \theta_k} - \sum_\textbf{b}\frac{P^\textbf{b}}{\sum_\textbf{c}P^\textbf{c}} \frac{\partial \log P^\textbf{b}}{\partial \theta_k}\\
=&\frac{\partial\log P^\textbf{a}}{\partial \theta_k} - \left\langle\frac{\partial \log P^\textbf{a}}{\partial \theta_k}\right\rangle_{\textbf{a}\sim P}.
\end{aligned}
\end{equation}
Here, the log derivative trick was used in the third line and we renamed the dummy indices $\textbf{b}$ and $\textbf{c}$ in the last step. One may proceed similarly for the time derivative. Overall, this leaves us with the connected correlator structure described in the main text
\begin{equation}
\begin{aligned}
S_{kk^\prime} &= \left\langle O^\textbf{a}_kO^\textbf{a}_{k^\prime} \right\rangle_{\textbf{a}\sim P} - \left\langle O^\textbf{a}_{k} \right\rangle_{\textbf{a}\sim P} \left\langle O^\textbf{a}_{k^\prime} \right\rangle_{\textbf{a}\sim P}\\
F_k &= \left\langle \mathcal{L}^\textbf{ab}\frac{P^\textbf{b}}{P^\textbf{a}} O^\textbf{a}_k \right\rangle_{\textbf{a}\sim P}-\left\langle O^\textbf{a}_k \right\rangle_{\textbf{a}\sim P}\left\langle \mathcal{L}^\textbf{ab}\frac{P^\textbf{b}}{P^\textbf{a}} \right\rangle_{\textbf{a}\sim P}.\\
\end{aligned}
\end{equation}

We finally arrive at
\begin{equation}
\dot{\theta}_k = \tilde{S}^{-1}_{kk^\prime}F_{k^\prime}
\end{equation}
where the tilde is due to the fact that we cannot invert $S$ directly but rather need to regularize it because it is usually ill-conditioned. One can easily show that the updates that were found are indeed maxima of the fidelity:

\begin{equation}
\begin{aligned}
&\frac{\partial^2}{\partial \dot{\theta}_k^2}F(P^a + \dot{P}^a \tau, P^a + \sum_{k^\prime} \frac{\partial P^a}{\partial \theta_{k^\prime}}\dot{\theta}_{k^\prime} \tau)\\
=&\frac{\partial}{\partial \dot{\theta}_k}(F_k - S_{kk^\prime}\dot{\theta}_{k^\prime})\\
=&-S_{kk^\prime}\delta_{k^\prime k}\\
=&-S_{kk}\\
=& - \left\langle \left(O^\textbf{a}_k-\left\langle O^\textbf{a}_k \right\rangle\right)^2 \right\rangle_{\textbf{a}\sim P}\\
\leq& 0.
\end{aligned}
\end{equation}

\section{Observables and Operators in the POVM-formalism}
As described in the main text, the POVM-distribution $P$ is obtained as expectation values of the respective POVM-operators $\hat M$,
\begin{equation}
P^\textbf{a} = \tr\left(\hat \rho \hat M^\textbf{a}\right) \,,
\end{equation}
where  $\hat M^\textbf{a}=\hat M^{a_1}\otimes .. \otimes \hat M^{a_N}$ are product operators. For IC-POVMs with the minimal number of $(d^2)^N$ elements, where $d$ is the local Hilbert space dimension ($d=2$ for spins), this relation can be inverted:
\begin{equation}
\hat \rho = P^\textbf{a}T^{-1 \textbf{a}\textbf{a}^\prime}\hat M^{\textbf{a}^\prime},
\label{eqn:rho_from_P_app}
\end{equation}
with the overlap matrix $T^{\textbf{aa}^\prime} = \tr\left(\hat M^\textbf{a}\hat M^{\textbf{a}^\prime}\right)$. We note that not every normalized probability distribution inserted into Eq.~\eqref{eqn:rho_from_P_app} results in a physical density matrix, as the positivity of $\hat \rho$ is not ensured.
The equation describing the dynamics of $P$ can be obtained from Eq.~$\eqref{eqn:rho_from_P_app}$ and the Lindblad master equation  according to
\begin{equation}
\dot{P}^\textbf{a} = \tr\left(\dot{\hat \rho} \hat M^\textbf{a}\right) = \mathcal{L}^\textbf{ab}P^\textbf{b}.
\end{equation}
The the part of the linear map $\mathcal{L}$ resulting from the von-Neumann term, i.e.\ the part accounting for the unitary evolution, is
\begin{equation}
\begin{array}{rl}
\dot{P}^{\textbf{a}}&=\tr\left(-i\left[\hat H,\hat  \rho\right] \hat M^\textbf{a}\right)\\
&=\tr\left(-i\left[\hat H, T^{-1\textbf{bb}^\prime}\hat M^{\textbf{b}^\prime}\right] \hat M^\textbf{a}\right)P^{\textbf{b}}\\
&=\tr\left(-i\hat H\left[T^{-1\textbf{bb}^\prime}\hat M^{\textbf{b}^\prime},\hat M^\textbf{a}\right]\right)P^{\textbf{b}}\\
&=U^{\textbf{ab}}P^\textbf{b},
\end{array}
\end{equation}
where the cyclicity of the trace was used in the last line. A similar expression can be found for the dissipative part
\begin{equation}
\begin{aligned}
D^\textbf{ab} = \gamma \tr \Bigl( \sum_i & \hat L^i T^{-1\textbf{bb}^\prime}\hat M^{\textbf{b}^\prime}\hat L^{i^\dagger}\hat M^\textbf{a}\\
&-\frac{1}{2}\hat L^{i^\dagger}\hat L^i \left\lbrace T^{-1\textbf{bb}^\prime}\hat M^{\textbf{b}^\prime},\hat  M^\textbf{a} \right\rbrace \Bigr),
\end{aligned}
\end{equation}
from which we set $\mathcal{L}$ together according to
\begin{equation}
\mathcal{L}^\textbf{ab} = U^\textbf{ab} + D^\textbf{ab}.
\end{equation}
The expectation value of any observable in physical index space may be correspondingly expressed in the POVM-formalism replacing $\langle \hat O \rangle = \tr\left(\hat \rho \hat O\right)$ by $\langle \hat O \rangle = P^\textbf{a}\Omega^\textbf{a}$. The numerical values of the coefficients $\Omega^\textbf{a}$ are obtained in similar fashion as the Lindbladian operator $\mathcal{L}$, namely by substituting $\hat \rho$ according to Eq.~(\ref{eqn:rho_from_P_app})
\begin{equation}
\langle \hat O \rangle = \tr\left(\hat \rho \hat O\right) = P^\textbf{a} T^{-1\textbf{a}\textbf{a}^\prime}\tr\left(\hat M^{\textbf{a}^\prime}\hat O\right)=P^\textbf{a}\Omega^\textbf{a}.
\end{equation}

\section{Details of the RNN-architecture}
As described in the main text, the RNN encodes the probability distribution $P^\textbf{a}$ as a product of conditionals, $P^\textbf{a}=\prod_i P(a_i|a_{<i})$. The formula implies that the network's knowledge of previous POVM outcomes $a_{<i}$ may alter the estimation of POVM outcome probabilities at site $i$. In the network architecture, this is ensured by passing a hidden state to the next lattice site where it enters the computation of the probability output. This hidden state may be regarded as a latent embedding of physical contextual information, and is required to accurately encode correlations in the physical system.
Our results are obtained using standard RNN cells which are known to have exponentially decaying correlation length \cite{Shen2019}. In scenarios, where this is expected to be insufficient, more advanced cells, such as the Long Short Term Memory (LSTM) \cite{Hochreiter1997}, whose correlation length decays algebraically \cite{Shen2019}, or the transformer \cite{vaswani2017attention} may be used instead. 

Since the RNN architecture was originally developed to tackle tasks associated with serial data, some changes are required in order to make it suitable for quantum applications. For one, to allow the treatment of 2D systems the RNN evaluation and sampling schemes need to be generalized.
Here, we adapt a scheme, introduced in \cite{HibatAllah2020}, that treats correlations along both spatial direction on equal footing.
Additionally, we enforce all symmetries present in the Lindbladian $\mathcal{L}$ by averaging all symmetry-invariant outcome configurations \cite{HibatAllah2020}. These include translational symmetries as well as point symmetries. We emphasize, that explicitly restoring these symmetries in our ansatz improved the accuracies of observables substantially.

Furthermore, we here lay out how the network is initialized. Product states, which form typical initial states in non-equilibrium time evolution, may be encoded to numerical precision in the network, by setting the biases of the output layer to the logarithm of the to be encoded 1-particle probability distribution while simultaneously setting all weights connecting to the output layer to zero. We may therefore attribute all accumulated error to imprecise updates during time-evolution and note that no preceding computations are required.

For our simulations we use RNNs implemented in the open source machine learning library JAX \cite{Bradbury2018}. An RNN is a generative model that works on sequential data, in which the bits of the sequential data are processed in an iterative fashion. The RNN fulfills two tasks: It assigns probabilities to a given POVM outcome configuration and, as a generative model, is capable of exact sampling, meaning that it can be programmed to output sample POVM configurations in agreement with the assigned probabilities. This is a major advantage of autoregressive networks compared to other network architectures, in which the sampling step is carried out using Markov Chain Monte Carlo schemes, which potentially may be plagued by long autocorrelation times. 

An RNN-cell is the basic building block of an RNN; RNN-cells may be stacked to form the complete RNN, increasing the representational power of the network. Let us first limit our considerations to RNNs with one layer, i.e. single RNN-cells. The input to every RNN-cell consists of two parts: For one, the physical POVM outcomes $\textbf{a}=a_1..a_N$ are fed into the model piece by piece. Here, each outcome is transformed to a one-hot encoded vector of length 4. Simultaneously, a hidden state of length $l$ which is initialized to zero, i.e. $\mathbf{h}_0=0$ is fed into the model. The first step consists of finding the first probability appearing in $P^\textbf{a}=\prod_i P(a_i|a_{<i})$, i.e. $P(a_1)$.
First, a new hidden state is computed
\begin{equation}
    \mathbf{h}_1 = \phi \left(W_h \cdot \mathbf{h}_0 + W_a \cdot \mathbf{a}_0 + \mathbf{b}_h\right).
    \label{eqn:RNN_cell}
\end{equation}
$\mathbf{a}_0$ is an input of length 4 carrying zeros, similar to the empty input $\mathbf{h}_0$ of length $l$. The parameters $W^{l(a)}$ consequently are matrices with shape $l\times l$ ($l\times 4$), while the bias vector $\mathbf{b}_h$ has length $l$. We choose the element-wise activation function $\phi$ to be the Exponential Linear Unit (ELU)
\begin{equation}
    \phi(x)=
    \left\{ \begin{array}{ll}
        x &x>0, \\
        \alpha\left(e^x-1\right) &x \leq 0.
    \end{array} \right.
\end{equation}
Two more sets of parameters $W_s$ ($\mathbf{b}_s$) with shape $4\times l$ ($4$) enter the computation of the output of the RNN-cell,
\begin{equation}
    P(a_1) = \sigma\left(W_s \cdot \mathbf{h}_1 + \mathbf{b}_s\right).
    \label{eqn:final_comp}
\end{equation}
Here $\sigma$ denotes the softmax-activation, 
\begin{equation}
    \sigma(\mathbf{x})_i = \frac{e^{x_i}}{\sum_i e^{x_i}}
\end{equation}
and the summation includes the four possible POVM-outcomes, allowing to interpret $P^{a_1}$ as a proper discrete probability distribution. Depending on the task at hand, one may either store the probability of a POVM outcome of interest or sample the first POVM outcome from $P^{a_1}$. 

Obtaining an expression for $P(a_2|a_1)$ is identical to the hitherto described procedure, by substituting $\textbf{h}_0$ for $\textbf{h}_1$ and $\textbf{a}_0$ for $\textbf{a}_1$ and, more generally, $\textbf{h}_i$ for $\textbf{h}_{i+1}$ and $\textbf{a}_i$ for $\textbf{a}_{i+1}$ in the following steps. Here, the `recurrent` nature becomes apparent, since the same parameters, i.e. the same network, is used in every computation step. 

If one desires to use deeper networks with $K$ layers, Eq.~(\ref{eqn:RNN_cell}) changes to 
\begin{equation}
    \mathbf{h}_i^k = \phi \left(W_h^k \cdot \mathbf{h}_{i-1}^k + W_a^k \cdot \mathbf{h}_i^{k-1} + \mathbf{b}_h^k\right).
    \label{eqn:hiddenstate_deep}
\end{equation}
$\mathbf{h}_i^k$ is then called the hidden state at layer $k$ at lattice site $i$. The computation of $P(a_i|a_{<i})$ is still analogous to Eq.~(\ref{eqn:final_comp}), i.e.
\begin{equation}
    P(a_i|a_{<i}) = \sigma\left(W_s \cdot \mathbf{h}_i^K + \mathbf{b}_s\right).
\end{equation}
As the product of these probabilities becomes exponentially small in the system size $N$, one stores the logarithm of the conditional probability instead of the probability itself. 

In two-dimensional systems, the situation is slightly more involved. One might be tempted to map the 2D system in a snake-like fashion to a one-dimensional system. However, using this method one observes that correlators of vertical neighbours are not encoded accurately as information may potentially get lost upon long traversing times in horizontal direction \cite{HibatAllah2020}. Instead, we opt to pass hidden states in a two-dimensional fashion, incorporating the dimensionality of the system as shown in Ref.~\cite{HibatAllah2020}. Herein, we once again change Eq.~(\ref{eqn:hiddenstate_deep}) to read
\begin{equation}
    \mathbf{h}_{ij}^k = \phi \left(W_h^k \cdot \mathbf{h}_{i-1j}^k + W_h^k \cdot \mathbf{h}_{ij-1}^k + W_a^k \cdot \mathbf{h}_{ij}^{k-1} + \mathbf{b}_h^k\right).
    \label{eqn:hiddenstate_deep_2D}
\end{equation}
This method can in principle be extended to three dimensional systems.

\section{Comparison to exact numerical simulations for small system sizes}
To obtain uncontroversial benchmarks, we test our method in system size regimes where exact dynamics is feasible. As benchmark systems we choose the 1D and 2D systems described in the main text in Fig.~\ref{fig:Carra_Comparison} and reduce the system size to $N=10$ spins in the 1D case and a $3\times 3$ lattice in the 2D case.
\begin{figure*}[t]
    \centering
    \includegraphics[width=\linewidth]{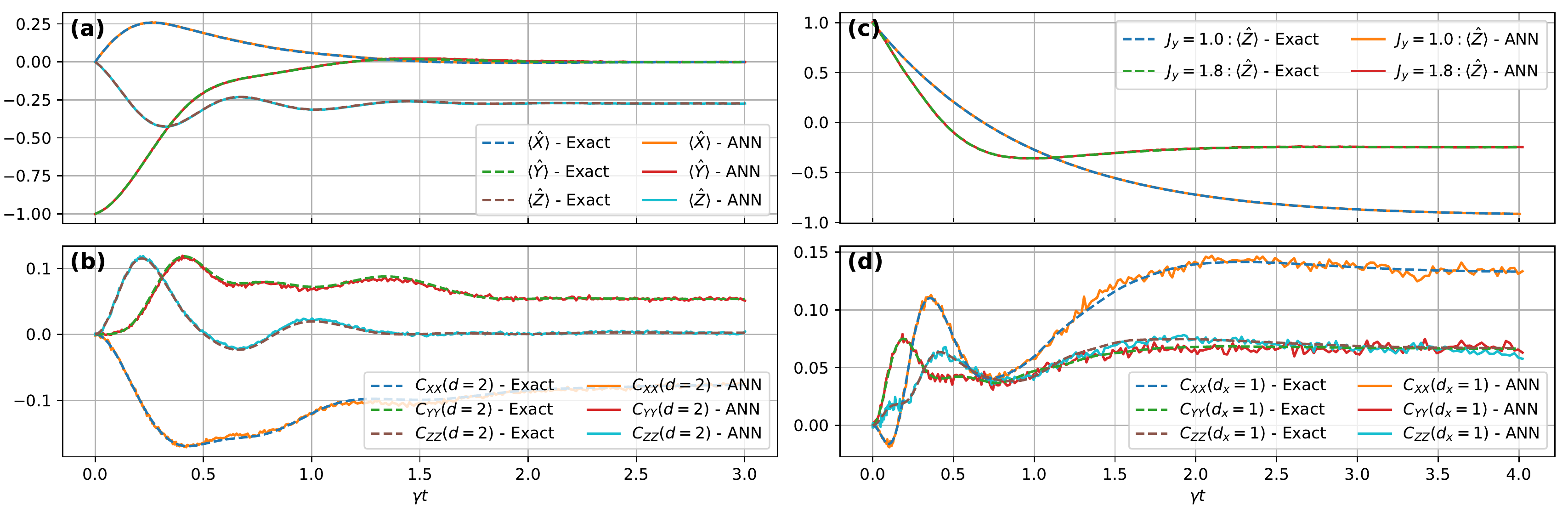}
    \caption{(a) and (b): Mean magnetizations and next-nearest neighbour connected correlation functions (e.g.\ $C_{XX}(d=2)=\sum_i\langle \hat X_i \hat X_{i+2}\rangle^c/N$) as a function of time in the anisotropic 1D Heisenberg model for $N=10$ spins starting in the product state $\langle \hat{Y}\rangle = -1$. Nearest neighbor couplings are given by $\vec{J}/\gamma =(2, 0, 1)$, $h_z/\gamma = 1$ and the dissipation channel is $\hat{L}=\hat{\sigma}^- =\frac{1}{2}(\hat{X}-i\hat{Y})$. The exact data is obtained for $N=10$ spins. (c) and (d): Mean $z$-magnetizations and nearest neighbour connected correlation functions (for $J_y/\gamma=1.8$) in a $3\times 3$ anisotropic 2D Heisenberg lattice with nearest neighbor couplings $\vec{J}/\gamma = (0.9, 1.0 (1.8), 1.0)$ and the same decay as in (a) and (b), starting in the product state $\langle \hat{Z} \rangle = 1$.}
    \label{fig:appx_Fig_2}
\end{figure*}

\section{Dissipative confinement correlations}
One question that arises in relation with Fig.~\ref{fig:Confinement} in the main text is how the spreading of correlations is to be described in the dissipative setting. As decoherence generically leads to classical transport dynamics one may expect diffusive growth of correlation that is proportional to $\sqrt{t}$ in contrast to the unitary linear light-cone proportional to $t$. 
\begin{figure}[ht!]
    \centering
    \includegraphics[width=\linewidth]{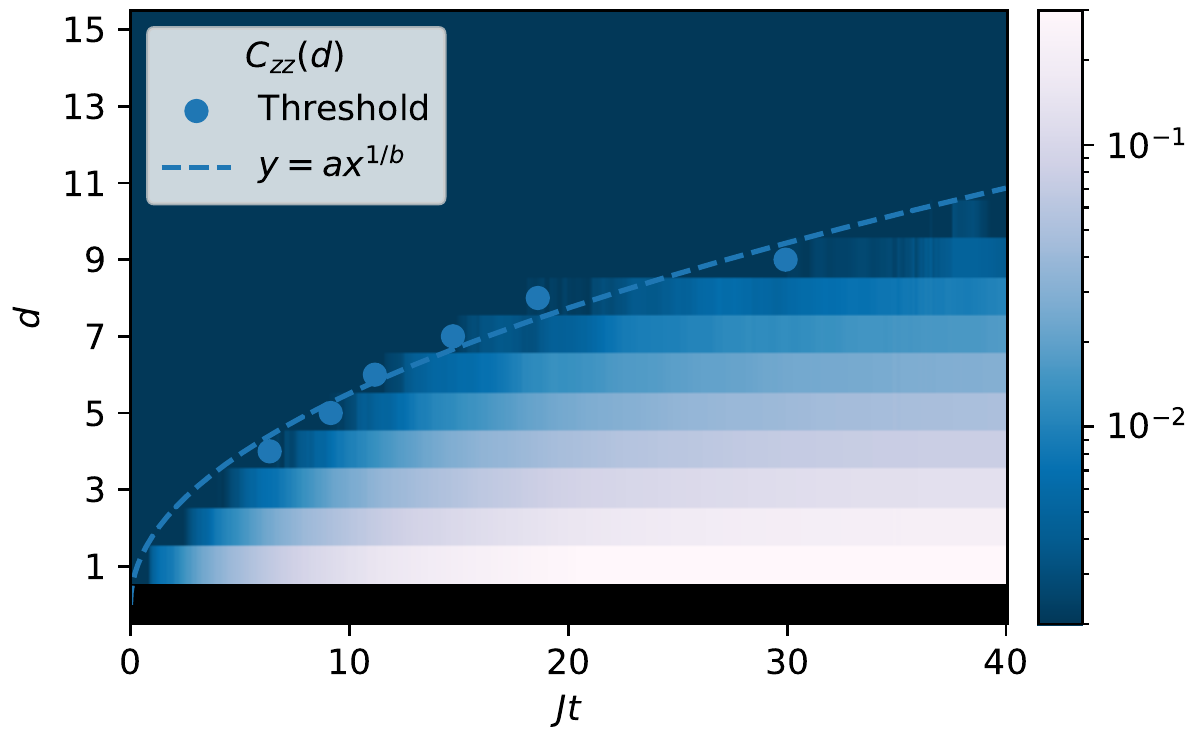}
    \caption{Spreading of correlations in the dissipative confinement model discussed in the main text (Fig.~\ref{fig:Confinement}). The data points are obtained as the first passages where $C_{ZZ}(d)\geq 0.002$ and the fitted curve $y=ax^{1/b}$ yields $b=2.04$.}
    \label{fig:sqrt_fit_appx}
\end{figure}
This intuition, however, can only hold in an intermediate time regime since at long times the system will relax to its steady state prohibiting an indefinite growth of correlations. 
In the present case of single particle dephasing noise the steady state is given by $\hat \rho(t\to\infty)\propto \mathds{1}$, as is easily verified by the observation that the unity operator commutes with $\hat H$ and similar for the dissipative part of the evolution Eq.~(\ref{eqn:Lindbladian_rho}) of the main text.
This means that all correlations will eventually decay to zero again in the long-time limit. One observes that correlations indeed start to disappear again at around $Jt\sim70$ in the considered setting.
Nevertheless, the spreading of correlations at intermediate times is consistent with a square-root, as shown in Fig.~\ref{fig:sqrt_fit_appx} where the dashed line is a fit to the first passage data points of a given threshold.

\begin{table*}[h]
    \centering
    \begin{tabular}{|c|c|c|c|c|c|c|}
    \hline
    Figure & Number of layers & Layer size & Number of parameters & Number of Samples & Integration tol. $\epsilon$\\\hline
    Fig. 2 (1D) & 3 & 20 & 2224 & 80.000 & 1e-05\\
    Fig. 2 (2D, $J_y/\gamma=1.0$) & 5 & 12 & 2224 & 8.000 & 1e-02\\
    Fig. 2 (2D, $J_y/\gamma=1.8$) & 3 & 20 & 3504 & 80.000 & 5e-03 \\
    Fig. 3 & 5 & 12 & 1456 & 160.000 & 1e-03 \\\hline
    \end{tabular}
    \caption{Hyperparameters that were used for the different figures in the main text. The integration tolerance is with respect to the $S$-matrix scheme proposed in \cite{Schmitt2020}.}
\end{table*}
\end{document}